\documentclass[traditabstract]{aa}
\usepackage{graphicx}
\usepackage{psfig}

\usepackage{txfonts}
\begin{document}
   \title{On the origin of the $\gamma$--ray emission from the flaring blazar PKS~1222+216}


   \author{F. Tavecchio\inst{1},
   J. Becerra-Gonzalez\inst{2,3},
   G. Ghisellini\inst{1}, A. Stamerra\inst{4}, G. Bonnoli\inst{1}, L. Foschini\inst{1}, L. Maraschi\inst{5}
          }

\institute{
   INAF -- Osservatorio Astronomico di Brera, via E. Bianchi 46, I--23807 Merate (LC), Italy\and
              Instituto de Astrofisica de Canarias (IAC), E-38200 La Laguna, Tenerife, Spain\and
              Depto. Astrofisica, Universidad de La Laguna (ULL), E-38206 La Laguna, Tenerife, Spain\and
              Dipartimento di Fisica, Universit\`a di Siena, and INFN Pisa, I-53100 Siena, Italy\and
              INAF -- Osservatorio Astronomico di Brera, via Brera 28, I--20121 Milano, Italy
             }

   \date{Received ; accepted }

 
  \abstract
{The flat spectrum radio quasar PKS 1222+216 (4C+21.35, $z=0.432$) was detected in the very 
high energy $\gamma$--ray band by MAGIC during a highly active $\gamma$--ray phase following an alert 
by the Large Area Telescope (LAT) onboard {\it Fermi}. 
Its relatively hard spectrum (70--400 GeV photon index $\Gamma=2.7\pm0.3$) 
without a cut off, together with the observed variability on timescale 
of $\sim 10$ min challenges standard emission models. 
In particular, if the emission originates in a portion of the relativistic jet located inside the 
broad line region (BLR), severe 
absorption of $\gamma$ rays above few tens of GeV is expected due to the $\gamma \gamma\rightarrow e^{\pm}$ process. 
These observations therefore imply the existence of a very compact ($R_{\rm b}\sim 5\times 10^{14}$ cm) and
very fast blob located far beyond the BLR radius (to avoid absorption), responsible for the rapidly 
varying high energy flux. However the long term (days-weeks) coherent evolution of the GeV flux recorded by LAT indicates that
there could be also the substantial contribution from another, larger, emission region. We model the spectral energy distribution of PKS 1222+216 during the epoch of the MAGIC detection assuming three different scenarios, namely: (1) a one-zone model considering only the emission from a compact blob outside the BLR;  
(2) a two-zone model considering the compact blob plus an emitting region encompassing the whole jet cross-section located outside the BLR and (3) a two zone model with the jet emitting region inside the BLR.
We find that in all cases the high-energy emission from the compact blob is dominated by the inverse Compton scattering of the IR thermal radiation of the dusty torus. Furthermore, both regions are matter dominated, with the Poynting flux providing a negligible contribution to the total jet power. These results disfavor models in which the compact blob is the result of reconnection events inside the jet or ``needles" of high-energy electrons accelerated close to the BH. Instead, the observational framework and our radiative models could be compatible with scenarios in which the jet is re-collimated and focussed at large distances from the central BH.

}{   }{   }{   }{    }

   \keywords{radiation mechanisms: non-thermal --- 
$\gamma$--rays: galaxies ---
quasars: individual: 4C+21.35       
}

\authorrunning{F. Tavecchio et al.}

\titlerunning{On the origin of gamma-ray emission of PKS 1222+216}

   \maketitle
%

\section{Introduction}

Blazars emitting $\gamma$ rays at Very High Energy (VHE, $E>100$ GeV) are still a minority of the 
whole population, but their number is steadily increasing due to the new generation of sensitive 
Cherenkov arrays (e.g. De Angelis et al. 2008, Hinton \& Hofmann 2009). 
The majority of them are BL Lac objects, in particular those with the synchrotron bump peaking 
in the UV/X--ray band (Highly peaked BL Lac, HBLs, Padovani \& Giommi 1995). 
The characteristic lack of important thermal components in their spectra suggests that the 
jet propagates in a relatively ``clean" environment. 
In this case it is widely believed that the high energy component is due to the 
inverse Compton (IC) scattering of the synchrotron photons themselves (synchrotron self--Compton model, SSC, 
e.g. Maraschi et al. 1992, Bloom \& Marscher 1996, Tavecchio et al. 1998).

Particularly interesting is the detection in the VHE band of few blazars belonging to the class of Flat 
Spectrum Radio Quasar (FSRQs).
The broad emission lines observed in their optical--UV spectra, typical of quasars, are though 
to originate in the so called broad line region (BLR), where  dense ($n\simeq 10^{9-10}$ cm$^{-3}$) 
clouds of gas orbiting at typical velocities of few thousands km/s at distances $R_{\rm BLR}\approx 10^{17-18}$ 
cm from the central black hole (BH) (e.g. Kaspi et al. 2007, Bentz et al. 2009) are photoionized by the UV 
continuum of the accretion disk. 
In this case {\it inside} the BLR, i.e. at distances $d<R_{\rm BLR}$,
the environment is rather rich of 
optical--UV photons, ideal targets for the IC scattering with the relativistic electrons in the jet. 
If the high--energy hump of FSRQs is dominated by this mechanism (External Compton scenarios, EC, e.g., Sikora et al. 1994, Ghisellini et al. 1998, Dermer et al. 2009) 
a pronounced softening of the spectrum is robustly expected above few tens of GeV due to the decreased 
scattering efficiency (e.g. Tavecchio \& Ghisellini 2008, Ghisellini \& Tavecchio 2009). 
More important, the dense radiation field makes the environment strongly opaque to $\gamma$ rays 
above few tens of GeV (e.g. Liu \& Bai 2006, Tavecchio \& Mazin 2009, Poutanen \& Stern 2010, but see Stern \& Poutanen 2011). 
The combination of the two mechanisms, effective above similar thresholds, 
makes the detection of VHE emission by FSRQs rather surprising.  
Even if the BLR is assumed to be ``flat", as suggested in the past (e.g. Shields 1978) 
and by recent studies (e.g. Jarvis \& McLure 2006, Decarli et al. 2011), the absorption cannot 
be avoided (Tavecchio et al., in prep). 

The prototypical FSRQ 3C 279 has been detected by MAGIC in two occasions, few days 
after optical flares (Albert et al. 2008, Aleksic et al. 2011a), with an extremely soft spectrum 
and a flux variable on a timescale of a day. 
PKS 1510--089 is another FSRQ detected at VHE by H.E.S.S. (Wagner \& Behera 2010).  
The recent detection of the FSRQ PKS 1222+216 (also known as 4C+21.35, $z=0.431$, Osterbrock \& Pogge 1987) by MAGIC 
(Aleksic et al. 2011b) is particularly challenging. 
The VHE spectrum smoothly connects to the LAT spectrum (Tanaka et al. 2011) and it is
well described by a rather hard power law, difficult to 
reconcile with the expected effects of absorption and decreased scattering efficiency discussed above. 
Moreover, the rapidly varying VHE flux, with a doubling of the flux in about 10 min, 
constrains the emitting region size to $R<c t_{\rm var} (1+z) 
\delta\simeq 2.5\times 10^{14} (\delta/10)$ cm ($\delta$ is the relativistic Doppler factor). 
If these dimensions are related to the jet cross sectional radius, as in the standard one--zone scenarios, 
this implies that the emission region is located close to black hole, 
hence deeply inside the opaque BLR radiation field.

As already proposed in Aleksic et al. (2011b), a possible way out to the problem is to 
assume that the region producing VHE $\gamma$ rays lies {\it beyond} the BLR, at 
distances $d>R_{\rm BLR}$. 
At these distances the external radiation field is likely 
dominated by the thermal radiation of the dusty torus reprocessing the disk radiation 
(e.g. B{\l}a{\.z}ejowski et al. 2000) and both the opacity and the reduced scattering 
efficiency problems are relaxed. However, even for extremely small values of the jet collimation angle ($<1$ deg), at 
these distances we expect that the jet cross sectional radius is far exceeding the limit imposed by the rapid variability 
(see also Tavecchio et al. 2010, Foschini et al. 2011). 
Hence we have to suppose  either (as proposed by Ghisellini \& Tavecchio 2008, 
Giannios et al. 2009, Marscher \& Jorstad 2010)
the existence of very compact emission regions embedded in the flow (as already done to explain the ultra-fast variability events observed 
in the two BL Lacs PKS 2155--304, Aharonian et al. 2007, and Mkn 501, Albert et al. 2007)  or strong recollimation and focusing of the flow (e.g. Stawarz et al. 2006, Bromberg \& Levinson 2009, Nalewajko \& Sikora 2009).

It is interesting to note here that the problems posed by the interpretation of these observational evidences 
from blazars have strict analogies with the issues raised by the unexpected recent discovery of fast 
($\sim $day) variability of the 
$\gamma$--ray emission of the Crab nebula by {\it AGILE} (Tavani et al. 2011) and {\it Fermi} (Abdo et al. 2011). In fact, as in blazars,
in order to account for these short timescales it seems necessary to admit the existence of very
compact knots inside the large scale relativistic wind from the pulsar (Vittorini et al. 2011, 
Bednarek \& Idek 2011) whose non-thermal emission is possibly boosted by relativistic effects 
(Komissarov \& Lyutikov 2011). It is therefore tempting to assume that
similar physical processes, possibly related to plasma instabilities in relativistic flows, operate 
and shape the emission in rather different $\gamma$--ray sources. 

In this paper we consider the problems posed by the observation of PKS 1222+216 focusing on the question if standard leptonic models can reproduce the observed spectral energy distribution (SED) and variability and, in particular, if a one-zone model is still a viable solution. To this aim we consider  
if a single, extremely compact emitting region (a ``blob") located beyond the BLR can account for the spectral and variability 
observed properties, concluding that, on the basis of the results of the SED modeling and on the power request, this solution could be viable. 
Alternatively, 
we study the scenario based on the existence of a two emitting regions, a ``standard" emission region encompassing the whole jet cross section plus 
a compact  ``blob" responsible for the rapidly varying high--energy emission detected by MAGIC. This possibility is in part motivated by the observed long-term smooth evolution shown by the LAT light-curve, which is difficult to explain if the emission originates from uncorrelated, fastly-evolving blobs. Moreover, we will show that this case is also energetically less demanding than the single blob model. 
This two-zone scenario resembles that already discussed by Ghisellini \& Tavecchio (2008) 
for the case of PKS 2155--304. However, the existence of an intense external radiation field makes the 
situation somewhat different, because the radiative interplay between the two regions is less important
than the EC emission with the IR photons (see Appendix B). 

We construct the SED using nearly simultaneous {\it Swift} (UVOT and XRT), {\it Fermi}/LAT and MAGIC data (\S 2). 
Then we apply the models to reproduce the SED (\S 3). Finally we discuss the results (\S 4).  

We adopt $H_{\rm 0}\rm =70\; km\; s^{-1}\; Mpc^{-1} $, $\Omega_{\Lambda}=0.7$, $\Omega_{\rm M} = 0.3$.

\section{The spectral energy distribution}

MAGIC detected PKS 1222+216 with a short ($\sim 30$ min) observation on 2010 June 17 (MJD 55364.9), during
a period of exceptional GeV activity (Tanaka et al. 2011). Data corrected for absorption by the interaction 
with the extragalactic background light modeled as in Dominguez et al. (2011) are reported in 
Fig. 1 (taken from Aleksic et al. 2011b). 
Aleksic et al. (2011b) also reported LAT data--points and the corresponding spectral ``bow--tie" 
obtained for a short time of 2.5 hours encompassing the MAGIC observation. 
For comparison, the black thick solid line (from Tanaka et al. 2011)  shows the LAT spectrum during quiescence.

{\it Swift} observed PKS 1222+216 several times in 2010 May--June, in correspondence with 
high activity of the source. 
Although there are no pointings exactly in time with the MAGIC observation, there is an 
observation few days after, on June 20, when the daily averaged LAT flux was similar to that on 
June 17 (see the LAT light curve in Tanaka et al. 2011).  

XRT (total exposure time of 4.5 ksec) data obtained during this observation (in PC mode) were analysed 
using the {\tt xrtpipeline} task considering grade 0--12 events and 
using v6.10 version of the {\tt HEASOFT} package and the calibration files {\tt caldb} 9/2/2011. 
As commonly observed in FSRQs, the XRT spectrum (red) is relatively hard with a 
photon index $\Gamma=1.60\pm0.13$ (the fit gives $\chi^2=9.25$ for 11 d.o.f.). 
The 0.3--10 keV flux is $F_X=5.5\times 10^{-12}$ erg cm$^{-2}$ s$^{-1}$. 
For comparison we also analyzed the data corresponding to a previous observation 
on May 29.  The flux was a factor $\approx 1.7$ higher, $F_X=8.3\times 10^{-12}$ erg cm$^{-2}$ s$^{-1}$, 
and the spectrum (cyan) was {\it softer} than on June 20, with $\Gamma=2.12\pm0.14$ ($\chi^2=10.6$ for 11 d.o.f.). 
It is interesting to note that while X-ray spectra of powerful FSRQs are generally hard, $\Gamma \leqslant 1.5$  (e.g. Kubo et al. 1998, Tavecchio et al. 2000, 2002), soft X--ray slopes (indicating a dominant contribution of the synchrotron or SSC component in this band) are occasionally shown also by the other two TeV detected FSRQs, 3C279 (e.g. Aleksic et al. 2011a) and 1510-089 (e.g. Tavecchio et al. 2000).

UVOT data--points (obtained with the standard procedure, e.g. Bonnoli et al. 2011) reveal a rather hard optical--UV spectrum suggestive of the presence of direct 
thermal emission from the accretion disc (see e.g. Ghisellini et al. 2009, 2011 for a discussion). 
To further investigate this point we also consider SDSS 
data
(taken on January 2008). 
To directly compare the UVOT and SDSS photometry we convert SDSS {\it ugriz} magnitudes into the standard 
Johnson system using the formulae provided by Chonis \& Gaskell (2008). 
Using standard zero--points we finally convert magnitudes into fluxes (magenta open squares). 
The SDSS data confirm the presence of a rather hard continuum. 
However, for the overlapping filters (U, B and V) there is a systematic difference of 
about 0.8 magnitudes between UVOT and SDSS, translating into UVOT fluxes larger by a 
factor of 2, as clearly visible in the SEDs. This difference could reveal an increase of the accretion 
luminosity between the SDSS (Jan. 2008) and the UVOT (June 2010) observations, possibly related 
to the high activity in $\gamma$ rays.

\begin{figure}
\vskip -0.5 cm
\hskip -0.3 cm
\psfig{file=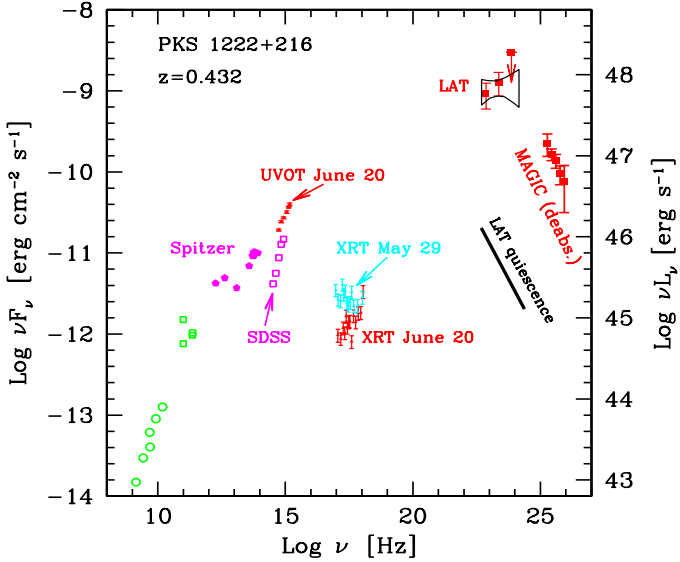,height=9.0cm,width=9.5cm}
\vskip -0.8 cm
\caption{
Spectral energy distribution of PKS 1222+216 close to the epoch of 
the MAGIC detection (2010 June 17). 
Red points at optical--UV and X--ray frequencies are from a {\it Swift} observation of June 20. For comparison,
cyan data-points show the X-ray spectrum two weeks before, on May 29 (see text).  
{\it Fermi}/LAT (red squares and ``bow tie") and MAGIC data (corrected for absorption by the EBL 
using the model of Dominguez et al. 2011) are taken from Aleksic et al. (2011b). 
The thick black solid line shows the LAT spectrum in quiescence (from Tanaka et al. 2011).
Magenta open squares are SDSS photometric points. Magenta filled pentagons are IR 
data from Malmrose et al. (2011). 
Green points report historical data (from NED, circles, and Tornikoski et al. 1996, squares). 
}
\label{sed}
\end{figure}

Malmrose et al. (2011) recently reported {\it Spitzer} observations in the IR band for four 
sources, including PKS 1222+216.
The IR data points (filled magenta pentagons in Fig.\ref{allblob}) track a 
bump around $3$ $\mu$m which is well fitted by a black body with temperature of 
$T=1200$ K, clearly related to the thermal emission from the putative dusty torus.
Finally, we also add historical radio (from NED) and millimeter (Tornikoski et al. 1996) data 
(green open circles and open squares, respectively).

\section{Modelling the SED}

\subsection{Observational facts and problems}
 
In modeling the observed SED we are constrained/guided by the following observational facts:

\noindent
1) The MAGIC VHE spectrum (70--400 GeV) is well described by a hard power law, with photon index (after correction for absorption by the interaction with the extragalactic background light) of $2.7\pm0.3$, and a cut-off for energies lower than 130 GeV is excluded. This spectrum smoothly connects with the LAT spectrum close to the MAGIC detection (Tanaka et al. 2011), strongly suggesting that high-energy and VHE emissions belong to a unique spectral component, originating in the same region.

\noindent
2) The MAGIC lightcurve shows a significative increase of the flux during the 30 min observation, with a doubling time of about $t_{\rm var}\simeq 10$ minutes. The causality relation $R<ct_{\rm var}(1+z)\delta$ allows us to constrain the size of the emitting region to $R<2.5\times 10^{14} (\delta/10)$ cm for typical values  of the Doppler factor $\delta=10$.

\noindent
3) The LAT long-term lightcurve (Tanaka et al. 2011) is characterized by periods of quiescence and smooth, long lasting ($\sim$ 1 week) flares. The MAGIC detection 
coincides with the raising part of a flare lasting for approximately 3 days. The $\gamma$--ray LAT flux ($F_{\rm >100 \,MeV}\sim 6.5\times 10^{-6}$ ph cm$^{-2}$ s$^{-1}$)  
was about half that recorded at the maximum of the flare ($F_{\rm >100 \,MeV}\sim 13.5\times 10^{-6}$ ph cm$^{-2}$ s$^{-1}$), reached the day after the MAGIC detection . 

Standard one-zone models for FSRQ generally assume that a single region in the jet, with a size comparable with that 
of the jet cross sectional radius, is responsible for the emission from IR to GeV frequencies. 
The location of this region is generally assumed to be inside the BLR (e.g. Dermer et al. 2009, Ghisellini \& Tavecchio 2009), 
but scenarios considering regions beyond it have been discussed (e.g., Sikora et al. 2008, Marscher et al. 2008).

The observational facts listed above already pose some problems to this view. Points 1) and 2) imply that the entire MeV-GeV and VHE emission component at the epoch of the MAGIC detection was produced in a very compact emission region outside the BLR, to minimize the expected severe absorption above 10 GeV (but see Stern \& Poutanen 2011). In the framework of one-zone models, 
a first possibility is therefore to assume that the entire $\gamma$--ray activity is due to the cumulative emission of  very compact, uncorrelated traveling regions (resulting from, e.g. internal shocks, Spada et al. 2001). However, in this case the expected {\it erratic} behavior is in contrast with  the smooth long-term evolution shown by LAT. One way to reconcile this scenario with point 3)  seems to assume the existence of a very compact and {\it stationary} region: 
this would allow fast variations of the flux and, at the same time, the long term modulation of the jet power would account for the smooth and coherent evolution. As an alternative we could envision the existence of {\it two} emitting regions, a large region responsible for the long-term evolution visible in the LAT band and an extremely compact region accounting for the fast variations.

Motivated by the arguments above, in the following we present {\it three} different scenarios for the VHE flare of PKS 1222+216 (see Fig. \ref{cartoon}). In the first case (A) we assume that the entire SED is produced by a single compact blob outside the BLR. In the other two cases we consider a two-zone model with the large region located outside (B) or inside (C) the BLR.  For consistency with the scenario sketched above, in cases B and C we admit that the large region could substantially contribute to (even if not dominate) the LAT band also at the epoch of the MAGIC detection. 

\subsection{Model setup}

\begin{figure}
\hspace*{-0.8cm}
\centerline{\psfig{file=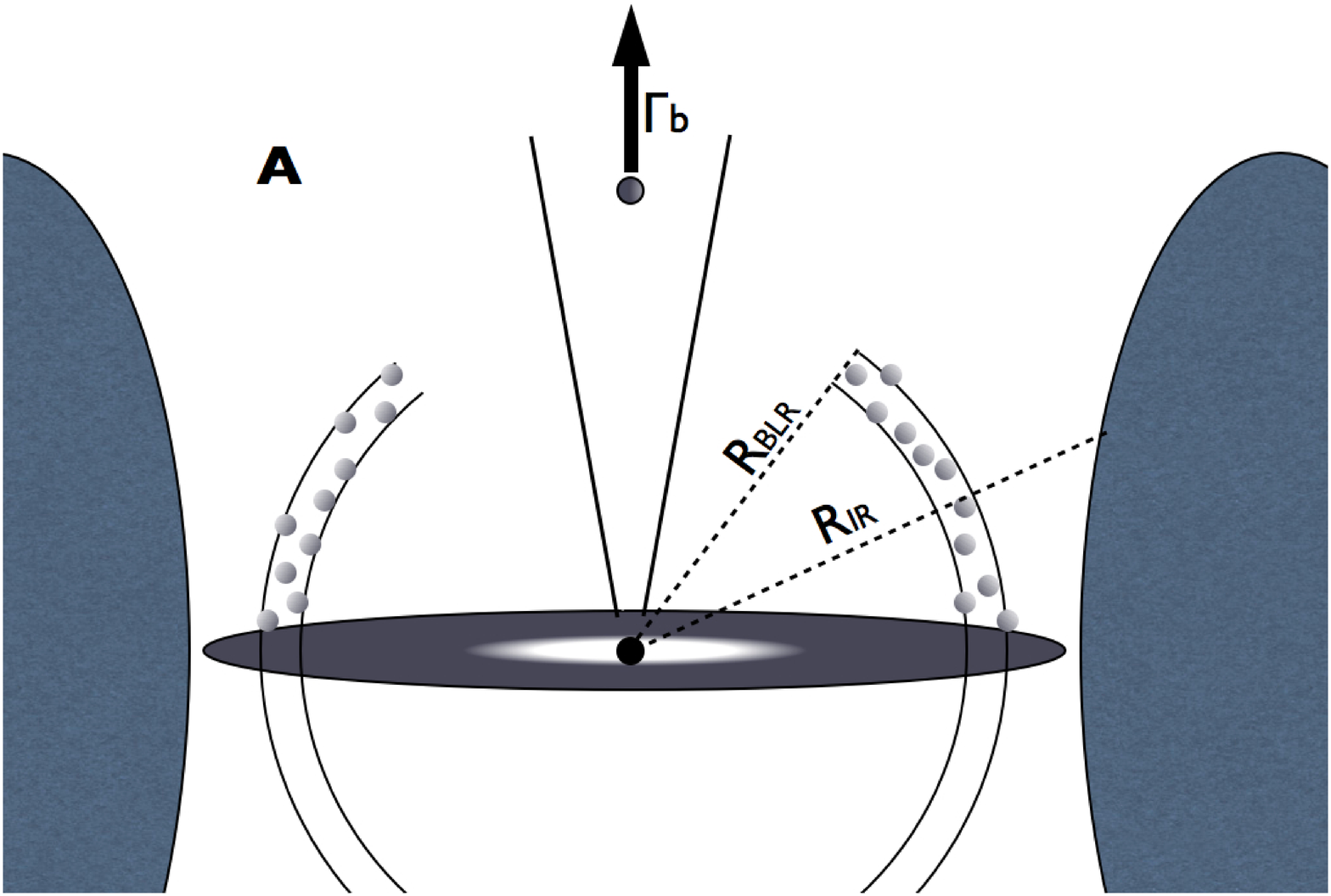,height=5.2cm}}
\centerline{\psfig{file=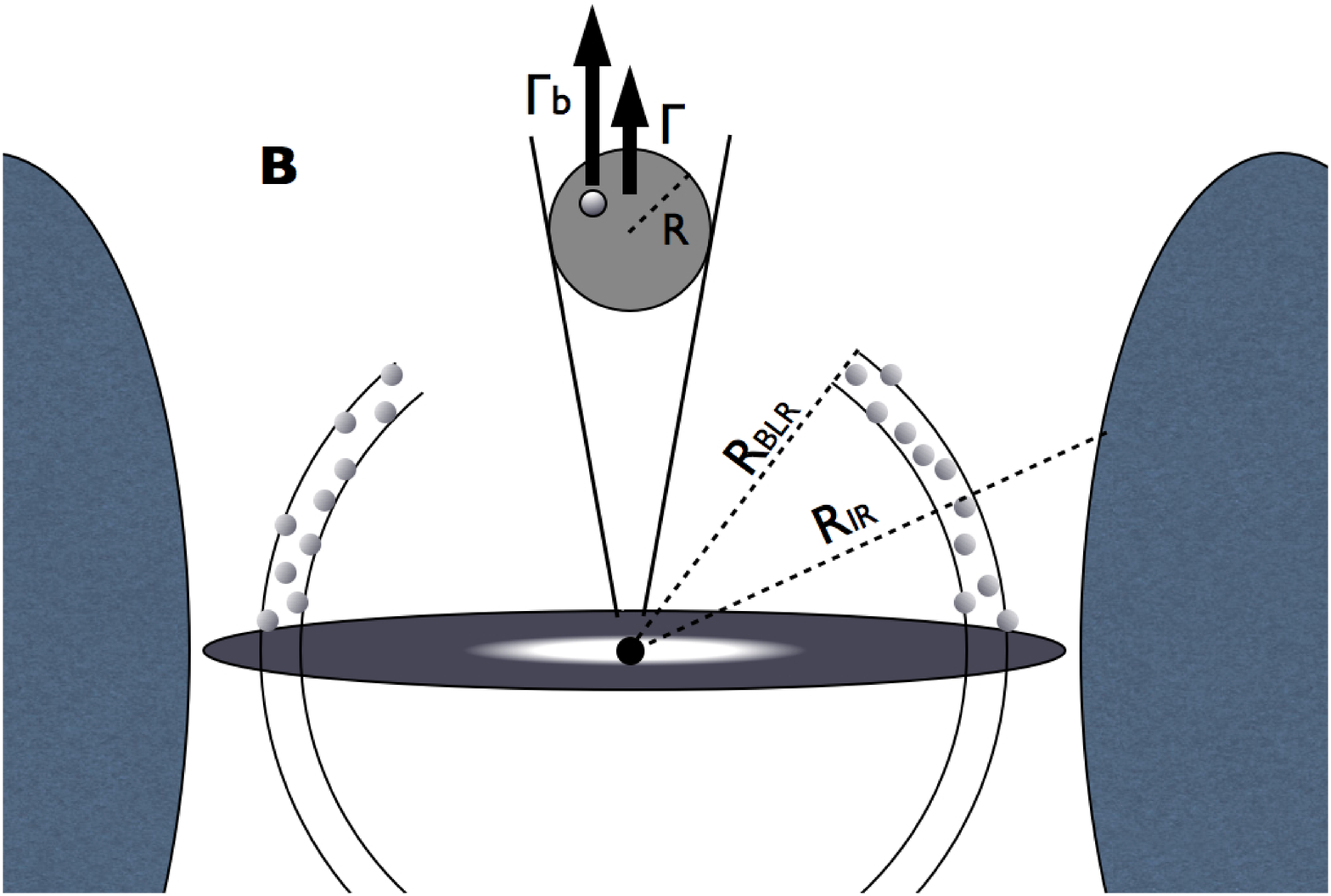,height=5.2cm}}
\centerline{\psfig{file=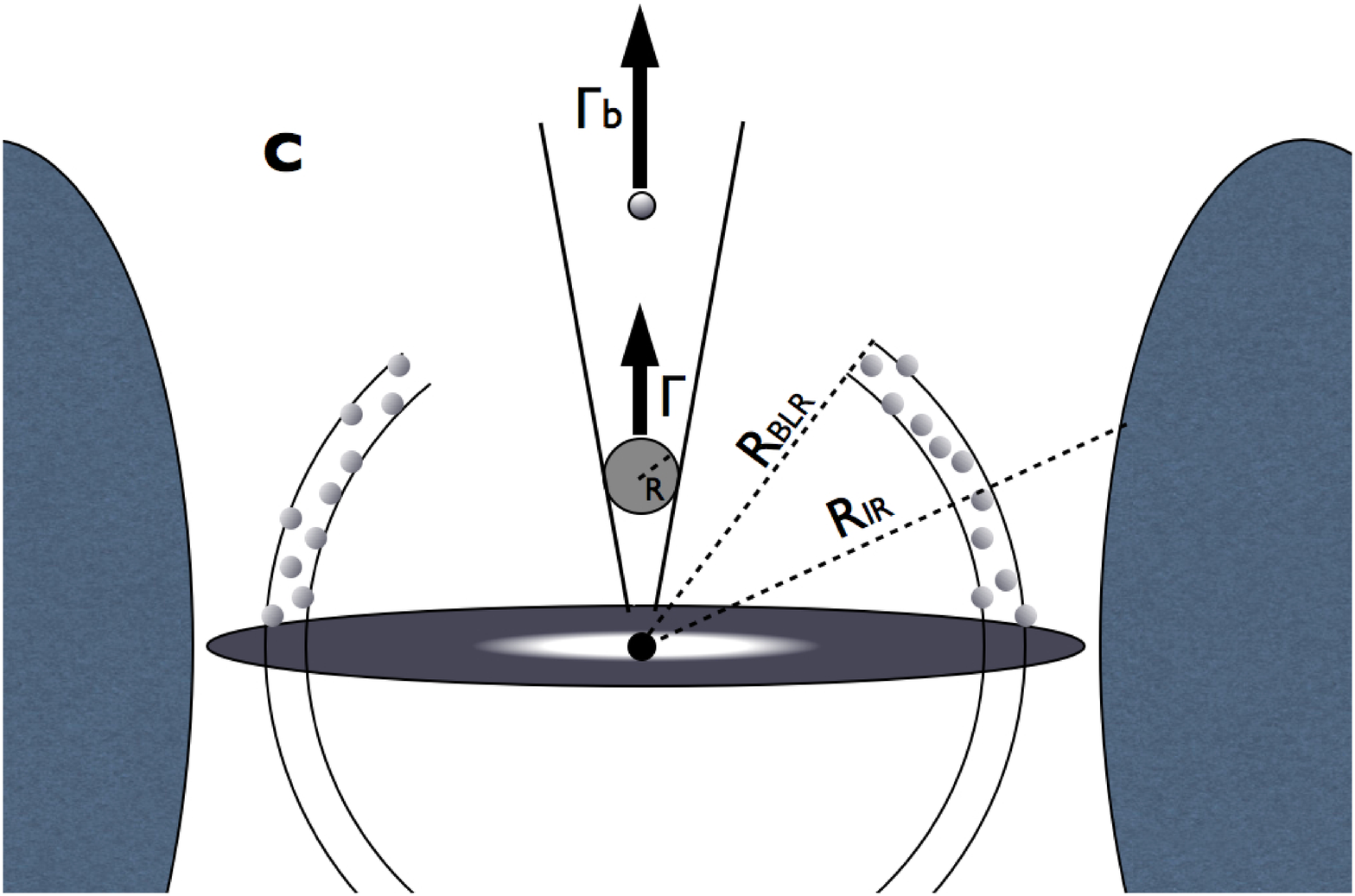,height=5.2cm}}
\vspace{0.3cm}
\caption{
Sketch of the geometrical arrangement assumed in the model (not to scale). We consider a spherical BLR with radius $R_{\rm BLR}$ and a dusty torus at $R_{\rm IR}$. 
In all the cases we consider the emission from a small compact ``blob" of radius $R_{\rm b}$ moving with Lorentz factor $\Gamma _{\rm b}$.
While in case A we suppose that the blob is responsible for the entire SED, in case B and C we also consider the emission from 
a ``standard" spherical emission with radius $R$ equal to the cross sectional size of a  conical jet with semi--aperture angle $\phi$, moving with bulk Lorentz factor $\Gamma$ located outside (B) or
inside (C) the BLR.
Each region is characterized by different values of the physical parameters, such as the 
magnetic field, electron density and energies. 
See text for more details.}
\label{cartoon}
\end{figure}

\begin{figure}
\vskip -0.5 cm
\hskip -0.3 cm
\psfig{file=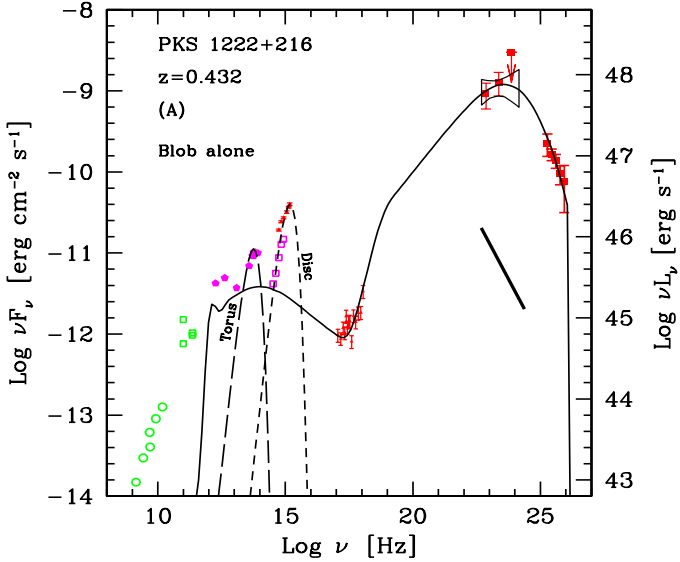,height=9.0cm,width=9.5cm}
\vskip -0.8 cm
\caption{
As in Fig. 1 but with the model considering only a compact emission region located 
beyond the BLR (see text for details). 
Black short and long dashed curve show the assumed emission from the accretion disk and dusty torus. For clarity  the XRT spectrum 
of May 29 is not reported.
}
\label{allblob}
\end{figure}

A sketch of the assumed geometry is shown in Fig. \ref{cartoon}.  In all cases a central BH is surrounded by an accretion disk  whose radiation, with luminosity $L_{\rm d}$,  photoionizes the BLR, modelled as a spherical shell located at distance $R_{\rm BLR}$ from the BH. Following Ghisellini \& Tavecchio (2009), we set 
$R_{\rm BLR}=10^{17} L_{\rm d, 45}^{0.5}$ cm. This relation provides a good approximation of the most recent results of the reverberation mapping 
studies (e.g. Kaspi et al. 2007, Bentz et al. 2009). We suppose that the BLR clouds intercept and reprocess 
(mainly into emission lines) a fraction $\xi_{\rm BLR}$ of $L_{\rm d}$. As discussed in Tavecchio \& Ghisellini (2008) a rather good approximation for the BLR radiation field as seen in the comoving frame is a black body peaked  at $\nu ^{\prime}_{\rm BLR}\approx 3\times 10^{15} \Gamma$ Hz.

Since the UVOT data--points probably trace the direct disk emission we fix $L_{\rm d}$ by reproducing 
the UVOT fluxes with a black body. 
Assuming that the peak is in correspondence with the UVOT filter at 
the highest frequency (UVW2), a lower limit for the disk luminosity is 
$L_{\rm d}=5\times 10^{46}$ erg s$^{-1}$ (Fig. 3, black short dashed line). 
This luminosity is exactly ten times larger than that estimated by 
Tanaka et al. (2011) by using the BLR total luminosity (and assuming $\xi_{\rm BLR}\sim 0.1$),  
in turn estimated by the luminosity of the H$\beta$ line in the SDSS spectrum (Fan et al. 2006). 
Considering the difference between the flux at the epoch of the SDSS (Jan 2008) 
and UVOT measures discussed in \S 2 the discrepancy is reduced by a factor of $\approx 2$. 
A difference by a factor of 5 between the two estimates could be explained by assuming $\xi_{\rm BLR}\sim 0.02$. Setting $L_{\rm d}=5\times 10^{46}$ erg s$^{-1}$, we have $R_{\rm BLR}=7\times 10^{17}$ cm.

Outside the BLR a dusty torus intercepts and re-emits part of the central disk emission.
The radiation field of the torus is modeled as a black body with temperature $T_{\rm IR}=1.2\times 10^3$ K with 
total luminosity $L_{\rm IR}=10^{46}$ erg s$^{-1}$  (black long dashed line on Fig. 3) filling a volume that,
for simplicity, is approximated as a spherical shell with radius $R_{\rm IR}=7\times 10^{18}$ cm.

In all the cases we model the compact emission region as a sphere with radius $R_{\rm b}$  (the subscript ``b" marks all the physical quantities related to the blob), moving with bulk Lorentz factor $\Gamma _{\rm b}$, filled with uniform and tangled magnetic field $B_{\rm b}$. 
We assume that relativistic electrons follow a smoothed broken power law energy distribution with 
normalization $K$ and slope $n_1$ and $n_2$ below and above the break at the energy $\gamma _{\rm p} m_e c^2$. 

We assume a conical jet propagating from the BH vicinity, with semi--aperture angle $\phi =0.1$ rad. A spherical region at distance $d$ and radius $R=\phi\, d$ 
moving along the jet with bulk Lorentz factor $\Gamma$, carries tangled magnetic field (with uniform intensity $B$) and relativistic electrons 
with a distribution of the same functional form as in the case of the blob.

In case (A) we consider only the emission from the compact blob. In case (B) and (C) we consider the emission from both regions. 

Both regions will emit synchrotron and IC radiation. 
For the blob and the jet components we consider as seed photons for the IC scattering the locally-produced synchrotron photons (SSC), the thermal 
emission from the dusty torus (EC/IR) and, in case (B), the synchrotron photons produced in the other 
region of the jet (we will call them the EC/J and EC/b component). In case (C) for the large region we consider also the photons reprocessed from the BLR (EC/BLR). 
The calculations have been performed adapting the code described in detail in Maraschi \& Tavecchio (2003).

Since by construction the contribution of the large region is always negligible above 20-30 GeV, for simplicity we 
neglect absorption for this component when considering the region inside the BLR. 
For the blob we expect a very weak absorption due to the
IR radiation field of the torus, starting to be important at energies higher than  
a few TeV. We also neglect this effect in the spectra shown in Figs. 3 and 4.

\subsection{Results}

We first consider (case A) the possibility that the very compact blob emitting TeV photons is responsible for the entire
SED, from IR up to TeV energies (see Fig. 3). 
As a consequence of the large compactness implied by the short variability timescale, the system naturally produces a powerful SSC component peaking in the X-ray band. As detailed in the Appendix A, the short variability timescale and the condition that the SSC component lies below the observed X-ray spectrum constrains the ratio $B_{\rm b}/\delta_{\rm b}^5\sim 10^{-9}$ G. Therefore either we adopt a ``standard" value of the magnetic field ($B _{\rm b}=0.1-1$ G) using extremely large Doppler factors, $\delta _{\rm b}>50$ or, conversely, we fix $\delta _{\rm b}$ to smaller values $\delta _{\rm b}\sim 10$ with extremely low magnetic fields, $B _{\rm b}\sim 10^{-4}$ G. In view of the possibility that the blob is the result of reconnection events in the jet, expected to produce rather fast outflows (Giannios et al. 2009), in our model we adopt the first possibility, using  $\delta _{\rm b}=75$.

As commonly found in FSRQs (e.g.  the discussion in Celotti \& Ghisellini 2008),  the condition that the the low energy tail of the EC component reproduces the X-ray spectrum implies that the value of $\gamma _{\rm min}$ is constrained to be close to unity.
As a consequence  the large number of particles at low energy pushes the 
total power demand of the region (calculated assuming one cold proton per electron) to extremely large 
values, close to $P_{\rm jet}=3\times 10^{47}$ erg s$^{-1}$ (see Table \ref{model}). 
It would be possible to decrease the requested power allowing larger values of 
the Doppler factor $\delta_{\rm b}$. However, even with an extreme value of $\delta_{\rm b}=100$ the power remains quite large. 
Such large jet powers are not uncommon among FSRQs in bright states (e.g. Ghisellini et al. 2010; see also Bonnoli et al. 2011 for the specific case of 3C454.3, which has a similar $L_{\rm d}$) and therefore we cannot exclude this possibility for PKS 1222+216.
 

In Fig. 4 we present the results for the two-zone model in the two cases discussed above.
In case B (left) we assume 
that both regions are cospatial and located outside the BLR, at distances $d>R_{\rm BLR}$; in case C (right) the large region is inside the BLR, while the blob is kept at the same distance as before. For the purposes of the calculation the exact distance of 
the two regions is not important, since the radiation field of the BLR and the torus are uniform within the corresponding radii.

As demonstrated in Appendix B, in case (B) we find that the radiative coupling between the two regions 
is less important than the EC flux using IR photons as seeds for the IC emission. Therefore the EC/J and EC/b components can be safely neglected.

Since, in both cases, we are dealing with two separate regions, the number of parameters is relatively large and therefore 
we have some freedom when selecting their values. However, the choice is not arbitrary: 
in the emission models shown in Fig. 4 we were guided by the following criteria. 

{\it Compact blob ---}
We assume that it dominates the $\gamma$--ray emission but it provides a small contribution in 
the X--ray band (but see below). 
This allows to fix $\gamma_{\rm p}$, to put a lower limit to $\gamma _{\rm max}$, 
and to impose a lower limit on the minimum Lorentz factor of the emitting electrons (thus decreasing the total power request). 
Similarly to case A (see discussion above) we assume a large bulk Lorentz factor to keep the SSC emission below the observed X-ray luminosity. Finally the magnetic field and the electron density are determined by the level of 
the $\gamma$--ray emission and by the condition that the synchrotron emission 
lies below the observed optical flux level. 

\begin{figure*}
\vspace{-0.5cm}
\psfig{file=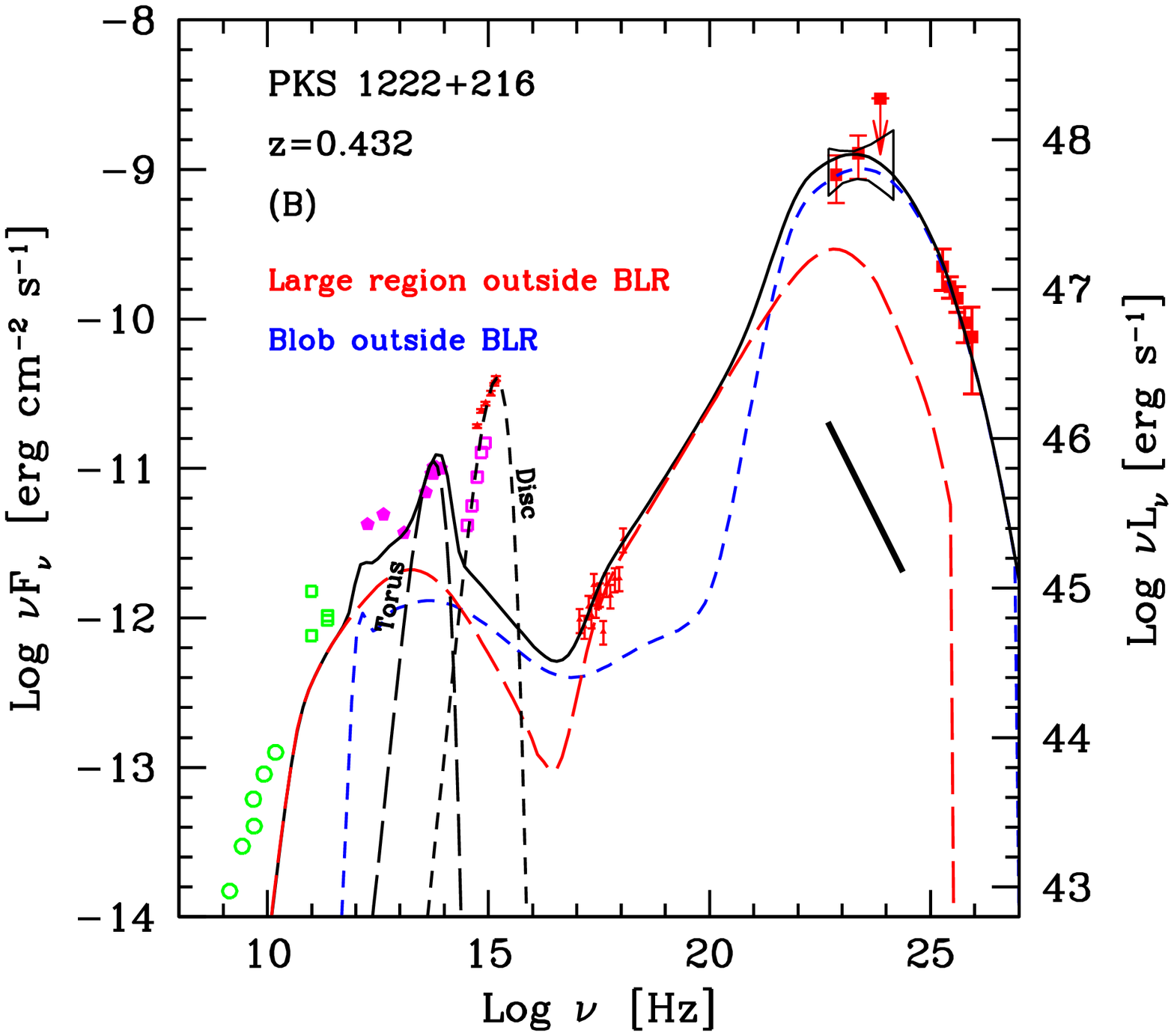,height=9.5cm,width=9.5cm}
\vspace{-9.5 cm}
\hspace*{9.1 cm}
\psfig{file=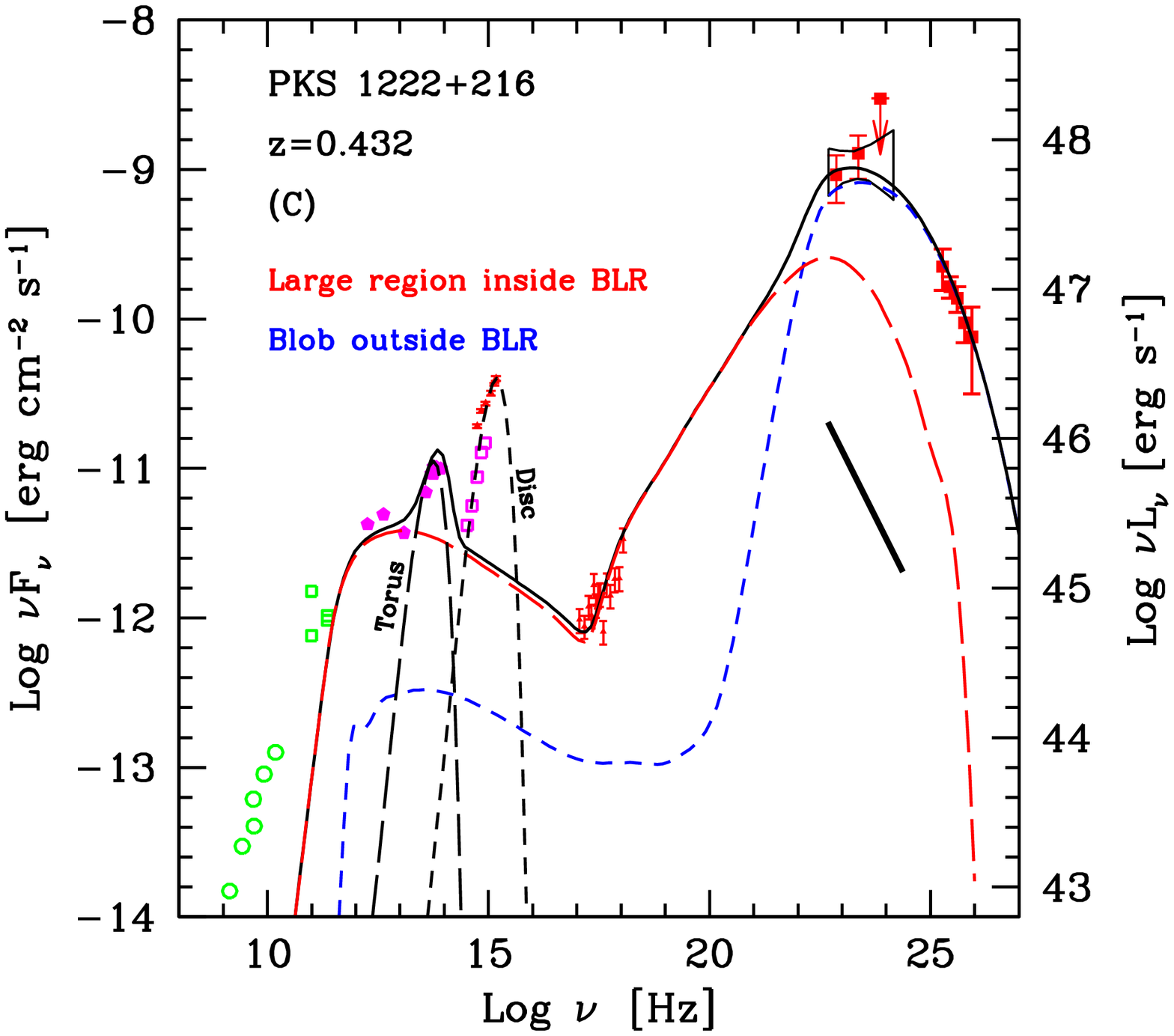,height=9.5cm,width=9.5cm}
\vspace{-0.8cm}
\caption{
As in Fig. \ref{allblob} but with the model corresponding to emission from the compact region and the standard jet. 
The blue short dashed line shows the emission from the compact region, while the red long dashed 
line reports the emission from the large region of the jet.  The solid black line is the sum of the two. 
The left panel shows the case in which both emission regions are located outside 
the BLR (case B); the right panel represents the case in which the large region is within the BLR (case C). 
}
\label{tworegion}
\end{figure*}

{\it Large region ---} For the IC component we are guided by the X--ray spectrum and by the observed flux. 
Moreover, we assume that its MeV-GeV emission contributes substantially to the total observed flux, although 
we assume that it lies a factor $\approx 2$ below the blob emission since the the LAT and the MAGIC spectra  suggest that they describe a unique component. The synchrotron peak is constrained by the new IR data.  
For case C (emission inside the BLR) we tried to use values of the physical parameters commonly inferred in FSRQs (e.g. Ghisellini et al. 2009). 

Concluding, while parameters cannot be uniquely determined, their values are 
reasonably constrained by the conditions posed by the SED and the observed variability. 
The parameters used  for the SEDs shown in Fig. 3-4 are reported in Table \ref{model}. 
Doppler factors are calculated assuming that the jet is observed at an angle $\theta _v=0.7^\circ$ from the axis.
This relatively small angle is required in order to allow the assumed large value of the Doppler factor of the blob, $\delta _{\rm b}=75$.
The last three columns report the power carried by cold protons, relativistic electrons and magnetic field. 
For the proton component we assume one proton per relativistic electron. 
For cases B and C the power carried by the jet is in both cases exceedingly larger than that of the blob. 
This is somewhat different than the case of similar models studied for the 
BL Lac PKS 2155--304 in Ghisellini \& Tavecchio (2008), where the power inferred for the blob 
was larger by more than one order of magnitude than that of the jet component. 
One reason for the difference is that the large external energy density makes the 
emission from the blob much more efficient than in  the case of PKS 2155--304 
(for which the dominant radiation field was that of the jet), thus reducing the 
required number of particles. Moreover, the total number of protons is also 
minimized by the fairly large minimum Lorentz factor of the emitting electrons. 
We also note that in all cases, both for the blob and the jet, the resulting intensity of the magnetic field is rather low. 
In turn this also implies that the associated Poynting flux is negligible compared to the power carried by the jet in the form of electrons and protons. The jet is therefore strongly matter dominated. This conclusion is robust, being the result of the the very large observed ratio between the high energy (IC) and low energy (synchrotron) bumps.

\section{Discussion and Conclusions}

The detection of intense, hard and rapidly variable VHE $\gamma$--ray emission from FSRQs, with 
thermal emission from the accretion disk and broad emission lines, poses a puzzle.
%
The observed very rapid variability constraints the size of the
emitting region to be small, suggesting that its location in the jet is inside the BLR.
But in this case the high energy photons interact with the emission line photons targets, and
get absorbed. 

The solution of this puzzle necessarily requires the existence of a compact region
located outside the BLR, whose size is smaller than the expected cross sectional radius of the jet at that
distance.
%
This scenario follows similar ideas proposed to explain the ultra--fast VHE emission observed 
in the BL Lac objects PKS 2155--304 and Mkn 501. 
However, the case of FSRQs is more constraining, since, to avoid the 
strong absorption by the BLR photons, the blob must be located at large distances, $\gtrsim 0.1$ pc. 
In BL Lacs, instead, the relatively ``clean" environment is rather transparent to 
$\gamma$--ray photons (at least up to the TeV band, above which the thermal radiation from a dusty 
torus could affect the emission, e.g. Celotti et al. 1998) and the emission region could reside much closer to the central engine.

As in the case of BL Lacs, also for FSRQs  such compact blobs could originate from reconnection 
events which could naturally produce compact regions of fastly moving plasma inside the jet 
(the so called ``jet in a jet" scenario, Giannios et al. 2009, 2010, Nalewajko et al. 2011). However, with respect 
to the case of BL Lac objects,
in which the emission and the electron cooling are dominated by the SSC process, 
the denser external radiation field is expected to dominate the radiation energy density, leading to 
stronger radiative losses, lower average particle energies and, 
consequently, an emission peaking at GeV energies. 
The derived total power carried by the ``blob" is much less than the total jet 
power, supporting the energetic sustainability of the process, and the possibility that there
are, at any given time, several active blobs, pointing in slightly different directions.
This would increase the otherwise small probability to have very small viewing angles
(see also discussions in Ghisellini et al. 2009; Giannios et al. 2010). 
It is conceivable that the events leading to the production of these fast blobs can occur 
also at smaller distances from the central BH, even within the BLR. 
In this case the dense radiation field would prevent the escape of the VHE 
photons and the entire power emitted above few tens of GeV would be 
eventually reprocessed at lower energies. 
A non--negligible magnetic field within the BLR would spread the directions
of the emitting pairs into a broad cone, thus decreasing the observed flux and 
the chances of detection of this reprocessed radiation.
Although this scenario is attractive, it is based on the assumption that the plasma in the emission region is magnetically dominated, with magnetization parameter (i.e. Poynting to kinetic flux ratio) of the order of 10--100. On the contrary, our models show that the flow at the distances where radiation is produced is characterized by a rather low value of the magnetic flux compared to that carried by electrons or protons.

\begin{table*}
\begin{center}  
\begin{tabular}{llllllllllllll} 
\hline  
   &$\gamma_{\rm min}$ &$\gamma_{\rm p}$ &$\gamma_{\rm max}$ &$n_1$ &$n_2$ &$B$ [G] &$K$ [cm$^{-3}$] &$R$ [cm] 
   &$\delta $ &$\Gamma$ &$P_{\rm p,45}$ &$P_{\rm e,45}$ &$P_{\rm B,45}$\\
\hline 
All Blob &3   &1200 &$5\times 10^4$ &2.2 &3.7 &0.28 &$3\times 10^6$    &$9.2\times 10^{14}$ &75   &50 &247  &2.2  &8.2$\times 10^{-4}$\\
\hline
Jet  (in)  &1   &300   &$6\times 10^4$ &2   &3.45 &0.75  &$1.8\times 10^4$ &$1.5\times 10^{16}$ &19.7 &10 &81 &0.34 &$6.3\times10^{-2}$\\
Blob &200 &700  &$4\times 10^5$ &2.2 &3.4 &0.1  &$6\times 10^6$    &$7.2\times 10^{14}$ &75   &50 &10.6 &0.4  &6.5$\times 10^{-5}$\\
\hline
Jet  (out) &3   &$2\times10^{3}$  &$6\times10^4$  &2   &4.1 &0.09 &$3\times 10^2$  &$1.1\times 10^{17}$ &19.7 &10 &17.7 &0.3  &  $5\times10^{-2}$              \\
Blob &100 &900  &$4\times 10^5$ &2.2 &3.6 &0.18  &$10^7$            &$6.2\times 10^{14}$ &75   &50 &3.9  &0.8  &1.5$\times 10^{-4}$\\
\hline\\
\end{tabular}
\end{center}
\vspace{-0.3cm}
\caption{
Input parameters for the emission models shown in Fig. 3 (first row) and Fig. 4 (other rows); 
``in" and ``out" means inside or outside the BLR.
The parameters are: the minimum ($\gamma_{\rm min}$), break ($\gamma_{\rm p}$) and maximum  ($\gamma_{\rm max}$) 
Lorentz factor and the low energy ($n_1$) and the high energy ($n_2$) slope of the smoothed power law electron 
energy distribution, the magnetic field $B$, the normalization of the electron distribution, $K$, the radius of 
the emission region, $R$, the Doppler factor $\delta$ and the corresponding bulk Lorentz factor $\Gamma$. 
Doppler factors are calculated assuming that the observer lies at an angle $\theta_{\rm v}=0.7^\circ$ from the jet axis. 
The last three columns report the power carried by (cold) protons (assuming one proton per emitting electron), 
relativistic electrons and magnetic field 
in units of $10^{45}$ erg s$^{-1}$.}
\label{model}
\end{table*}

The severe radiative losses of the electrons in the disk and BLR radiation fields
seem to rule out also the alternative possibility that the rapidly varying VHE emission originates in extremely collimated beams 
of ultra--relativistic electron, as envisaged in the ``needle" scenario of Ghisellini et al. (2009). 
In fact, due to the strong dependence of the
maximum electron Lorentz factor resulting from the competition between the magneto--centrifugal acceleration 
in the black hole magnetosphere (e.g., Rieger \& Mannheim 2000) and the strong IC cooling on the radiation 
energy density, $\gamma _{\rm max}\propto U_{\rm rad}^{-2}$ (Osmanov et al. 2007), the presence of a luminous 
disk and BLR will result in rather low energy of the electrons, emitting at energies below the GeV band.

Another possibility to explain the existence of a very small region at large distance is by the focusing of the jet through strong recollimation (e.g. Komissarov \& Falle 1997, Sokolov et al. 2004 Stawarz et al. 2006, Nalewajko \& Sikora 2008, Bromberg \& Levinson 2009). In particular, Bromberg \& Levinson (2009) showed that if the shock resulting from the recollimation of the jet by the interaction with an external medium is efficient in converting the dissipated energy into radiation, the reduced post-shock pressure allows the jet to be strongly ``squeezed", reaching a minimum size close to the point where the recollimation shock crosses the jet axis. At this point reflection (internal) shocks are expected to form (e.g., Komissarov \& Falle 1997), and the shortest variability time should be constrained by the size of these structures, of the order of the radius of the ``nozzle" of the jet. 
Under the assumption that the external confining medium is associated with matter in the BLR,  Bromberg \& Levinson (2009) estimate that the re-collimation shock reaches the jet axis at a distance $d^{\bigstar}$ of the order of $d^{\bigstar}\simeq 2.5 \, L_{\rm j,46} \, (R_{\rm BLR}/0.1 \, {\rm pc})^{-1}$ pc where $L_{\rm j}$ is the jet power. Their simulations show that at this point if ({\it i}) the cooling operates at a rate larger than that of the dynamical timescale and ({\it ii}) the radiative conversion efficiency is sufficiently large (30 percents of the bulk luminosity converted into radiation), the cross sectional radius of the jet in the focusing point can be as small as $R\sim 10^{-2.5} d^{\bigstar}$. Condition ({\it i}) is easily fulfilled in FSRQs, even outside the BLR. Nalewajko \& Sikora (2009) showed that the efficiency of internal energy production in recollimation shocks can be larger than 30\% provided the product of the jet bulk Lorentz factor and the aperture angle is larger than $\approx 3$. Therefore, provided that the further conversion of this dissipated energy into radiation can reach  large efficiencies, also condition ({\it ii}) could be satisfied. Considering the rough consistency of these numbers with the results of our radiative models and the difficulties faced by the ``jet in a jet" or ``needles" models discussed above, we believe that such a scenario is rather promising and worth to be better investigated.

\begin{acknowledgements}
We are grateful to an anonymous referee for rather constructive comments that leaded to substantially improve the paper. 
We thank Amir Levinson for discussions.
\end{acknowledgements}

\appendix

\section{Limit to $B$ and $\delta$ from the constraints on the SSC component}

One can derive an approximate relation between the magnetic field and the Doppler factor (assuming for simplicity $\delta_{\rm b}=\Gamma _{\rm b}$) starting from the approximate expressions for the EC and SSC luminosities:
\begin{equation}
L_{\rm SSC}=\frac{4}{3}\sigma _{\rm T} c U^{\prime}_{\rm synch} <\gamma ^2> n V \delta _{\rm b}^4
\end{equation}
\noindent
and:  
\begin{equation}
L_{\rm EC}=\frac{4}{3}\sigma _{\rm T} c U_{\rm ext} \Gamma _{\rm b}^2 <\gamma ^2> n V \delta _{\rm b}^4
\end{equation}
\noindent 
where $U^{\prime}_{\rm synch}$ and $U_{\rm ext}\Gamma_b^2$ are the energy density of the synchrotron and external radiation in the blob frame, $V=4/3\pi R_b^3$ is the blob volume, $n$ is the electron density and $<\gamma ^2>$ is the averaged value of the squared of the electron Lorentz factor.

Using these equations imposing that $L_{\rm EC}=L_{\gamma}$, $L_{\rm SSC}/L_{\rm EC}<L_{\rm X}/L_{\gamma}$ and $R_{\rm b}\simeq c t_{\rm var} \delta _{\rm b}$, with some manipulations one can derive the following relation between the magnetic field and the Doppler factor (assuming for simplicity $\delta_{\rm b}=\Gamma _{\rm b}$) containing only observed quantities:
\begin{equation}
B_{\rm b}\lesssim \left[ \frac{32\pi^2c^3}{L_{\gamma}} \left(  \frac{L_{\rm X}}{L_{\gamma}}\right)   \right]^{1/2} U_{\rm ext} \, t_{\rm var} \, \delta _{\rm b}^5.
\end{equation}
In our specific case with $t_{\rm var}=600$ s, $L_{\gamma}\simeq 3\times 10^{47}$ erg s$^{-1}$, $L_{\rm X}/L_{\gamma}\lesssim 10^{-3}$ and $U_{\rm ext}\equiv U_{\rm IR}=5\times 10^{-4}$ erg cm$^{-3}$ we derive $B_{\rm b}\simeq 10^{-4} (\delta _{\rm b}/10)^5$ G.

\section{Importance of the blob/jet radiative interplay}

The ratio of the EC/IR and EC/J (inverse Compton produced in the blob scattering photons of the jet) luminosities is given by the ratio of the energy densities of the corresponding seed photons in the blob rest frame:
\begin{equation}
\frac{L_{\rm EC/IR}}{L_{\rm EC/J}} = \frac{ U^{\prime}_{\rm IR}  }{U^{\prime}_{\rm J}}
\end{equation}
\noindent
where primed quantities are measured in the blob frame. The two energy densities can be expressed as:
\begin{equation}
U^{\prime}_{\rm IR} = \frac{L_{\rm IR} \Gamma_{\rm b}^2}{4\pi R_{\rm IR}^2 c}; \;\;\;\;\;  U^{\prime}_{\rm J} = \frac{L_{\rm J}\Gamma_{\rm rel}^2}{4\pi R_{\rm J}^2  \delta_{\rm J}^{4}c}
\end{equation}
\noindent
where ``b" and "J" refers to quantities related to the blob and the jet, respectively, and $\Gamma_{\rm rel}=\Gamma_{\rm b}\Gamma_{\rm J}(1-\beta _{\rm b}\beta _{\rm J})$ is the relative bulk Lorentz factor between the jet flow and the blob. Therefore we finally find:
\begin{equation}
\frac{L_{\rm EC/IR}}{L_{\rm EC/J}} = \frac{L_{\rm IR}}{L_{\rm J}} \left( \frac{R_{\rm J}}{R_{\rm IR}} \right)^2 \left( \frac{ \Gamma_{\rm b}}{\Gamma _{\rm rel}}\right)^2 \, \delta_{\rm J}^4.
\end{equation}
Since $L_{\rm IR}$ is comparable to the observed $L_{\rm J}$ (see Fig. 1), $R_{\rm IR} \sim 10 \times R_{\rm J}$, $\Gamma _{\rm b}=50$, $\Gamma _{\rm J}=10$, $\delta _{\rm J}\sim 20$, the ratio is largely in favor of the EC/IR component, $L_{\rm EC/IR}/L_{\rm EC/J}\approx 10^6$. 
\\

Analogously, one can evaluate the ratio of the EC/IR and EC/b (inverse Compton produced in the jet scattering photons of the blob) luminosities:
\begin{equation}
\frac{L_{\rm EC/IR}}{L_{\rm EC/b}} = \frac{ U^{\prime}_{\rm IR}  }{U^{\prime}_{\rm b}}
\end{equation}
where now primed quantities are measured in the jet frame. As above, this expression can be rewritten as:
\begin{equation}
\frac{L_{\rm EC/IR}}{L_{\rm EC/b}} = \frac{L_{\rm IR}}{L_{\rm b}} \left( \frac{R_{\rm J}}{R_{\rm IR}} \right)^2 \left( \frac{ \delta_{\rm b}}{\delta _{\rm rel}}\right)^4 \, \frac{1}{\Gamma_{\rm J}^2}.
\end{equation}
Again, since $L_{\rm IR}$ is comparable to or larger than the observed $L_{\rm B}$, $R_{\rm IR} \sim 10 \times R_{\rm J}$, $\Gamma _{\rm b}=50$, $\Gamma _{\rm J}=10$, $\delta _{\rm J}\sim 20$, $L_{\rm EC/IR}/L_{\rm EC/b}>> 1$.

Therefore both EC/J and EC/b components can be safely neglected.

\end{document}